\documentclass[aps,prb,twocolumn,groupedaddress]{revtex4-1}

\usepackage{graphicx}

\usepackage{amssymb}
\usepackage{amsmath}

\begin{document}

\title{Calculating the electromagnetic characteristics of bifacial optical nanomaterials}

\author{P. Grahn}
\author{A. Shevchenko}
\author{M. Kaivola}

\affiliation{Department of Applied Physics, Aalto University, P.O. Box 13500, FI-00076 Aalto, Finland}

\date{\today}

\begin{abstract}
We introduce a formalism that describes the interaction of light with bifacial optical nanomaterials. They are artificial noncentrosymmetric materials in which counter-propagating waves behave differently. We derive electromagnetic material parameters for uniaxial crystalline media in terms of the complex transmission and reflection coefficients of a single layer of the constituent nanoscatterers, which makes the numerical evaluation of these parameters very efficient. In addition, we present generalized Fresnel coefficients for such bifacial nanomaterials and investigate the fundamental role of higher-order electromagnetic multipoles on the bifaciality. We find that two counter-propagating waves in the material must experience the same refractive index, but they can have dramatically different wave impedances. The use of our model in practice is demonstrated with a particular example of a bifacial nanomaterial that exhibits a directional impedance matching to the surrounding medium.
\end{abstract}

\maketitle

\section{Introduction}

Optical nanomaterials are artificial substances with subwavelength-sized unit cells that contain specially designed nanoparticles (artificial atoms). Light propagation in such a material is fully determined by the way electromagnetic waves are scattered and absorbed by the nanoparticles. The optical characteristics of the material can be described using effective material parameters, such as refractive index $n$ and wave impedance $\eta$ when the nanomaterial can be treated as homogeneous. For a given nanomaterial design, these parameters are commonly retrieved by calculating the light transmission and reflection by a slab of such nanomaterial.\cite{Menzel08,Smith02,Menzel08-2} Using these retrieval procedures, certain nanomaterials have been designed to exhibit extreme values for the material parameters, which cannot be found in nature.\cite{Pshenay09,Choi11,Silveirinha07,Yuan07,Rockstuhl09,Paul09,Li13} The commonly used retrieval procedures rely upon the assumption that the nanomaterial is free of spatial dispersion and, thus, treatable in terms of standard Fresnel transmission and reflection coefficients. These assumptions, however, may not hold for materials composed of asymmetric nanoscatterers, such as typical split-ring-resonators used to obtain artificial magnetism.\cite{Gompf2012}

Nanomaterials in which two counter-propagating optical waves see the medium differently belong to a specific class of optically bifacial nanomaterials. The nanoscatterers in these materials are not centrosymmetric. In this work, we consider uniaxial noncentrosymmetric nanomaterials. More complicated nanomaterials, such as those composed of chiral nanoparticles, are left out of the scope of the present paper. Although asymmetric scatterers have been widely studied in terms of their scattering properties\cite{MD2,Pakizeh09}, magnetic near-field enhancement\cite{Pakizeh2008,Pakizeh12}, localized absorption\cite{Anto12}, color switching\cite{Shegai11,AlaviLavasani12} and mimicking electromagnetically induced transparency\cite{Liu09,Bozhevolnyi2011,Liu11}, no adequate theory currently exists for the description of nanomaterials constructed of such scatterers. For example, for a bifacial nanomaterial, the standard Fresnel coefficients cannot be applied. In this paper, we develop the necessary formalism - that allows calculation of the material parameters $n$ and $\eta$ - and use it to investigate the key properties of such artificial bifacial media. In particular, we find that two counter-propagating waves in the material must experience the same refractive index, but they can have dramatically different wave impedances.

We apply our model to a particular example of optical nanomaterials consisting of metal nanodimers and verify the correctness of our model by rigorous numerical calculations. We also verify that in the limiting case of symmetric nanoscatterers, our model is in agreement with the results of the existing retrieval procedures.

\section{Theory}
\subsection{Effective material parameters}\label{SecMP}

\begin{figure}
\includegraphics[scale=1]{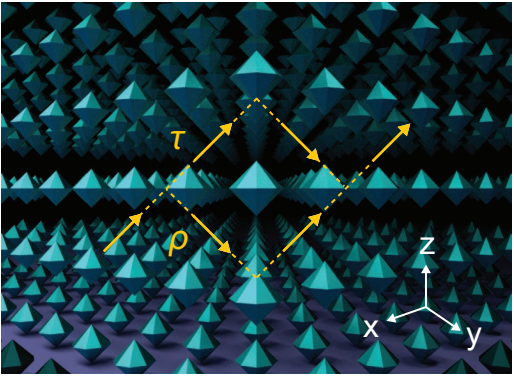}
\centering
\caption{Illustration of a three-dimensional nanomaterial composed of periodically arranged nanoscatterers. Light propagating in the nanomaterial can be described in terms of plane waves reflected back and forth by successive crystal planes. Each such plane is characterized by a transmission coefficient $\tau$ and reflection coefficient $\rho$. \label{figC}}
\end{figure}

\begin{figure}
\includegraphics[scale=0.8]{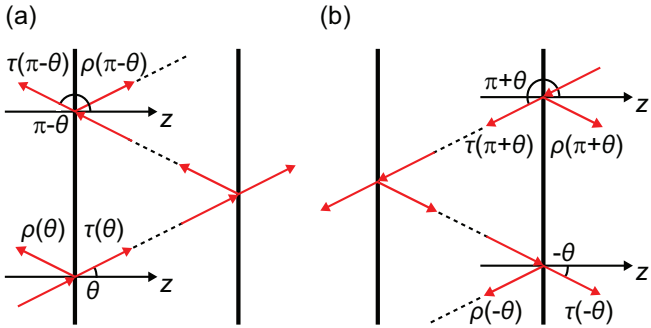}
\centering
\caption{Illustration of transmission and reflection of a plane wave propagating inside a nanomaterial. The vertical lines represent planes of nanoscatterers within the nanomaterial and the red arrows stand for the wave vectors. In (a) and (b) the field propagates between the planes at an angle of $\theta$ and $\pi+\theta$, respectively. \label{fig1}}
\end{figure}

We start by considering a three-dimensional nanomaterial that consists of periodically arranged nanoscatterers in a transparent dielectric host medium of refractive index $n_\mathrm{s}$. This three-dimensional array can be thought as a set of two-dimensional nanoscatterer arrays, which are stacked in a certain common direction, as illustrated in Fig.~\ref{figC}. We choose the $z$ axis to point in this direction and the length of the unit cell along $z$ to be $\Lambda_z$. For an optical plane wave propagating in the material, each two-dimensional nanoscatterer array can be treated as an infinitesimally thin sheet with a plane-wave transmission coefficient $\tau$ and reflection coefficient $\rho$, which in general both depend on the wave propagation direction and polarization. As has been shown in Ref.~\onlinecite{Interfero}, these coefficients can be obtained by calculating the plane-wave transmission and reflection by an isolated two-dimensional nanoscatterer array in the host medium. As long as the periodicities within the array are sufficiently small compared to the local wavelength $\lambda$, there will be no coupling of diffraction orders between successive ``crystal" planes in the material. Moreover, in the presence of sufficiently large spacing between neighboring nanoscatterers, the evanescent-wave coupling between the crystal planes can be safely neglected.\cite{Interfero,Alaee13,PA12} Thus, in the analysis, there will locally be only two counter-propagating plane waves in the crystal, due to the finite reflection provided by each crystal plane. The optical response of the nanomaterial to a plane wave propagating \emph{in the host medium} at an angle $\theta$ with respect to the $z$ axis is then fully determined by $\tau(\theta)$, $\tau(\pi-\theta)$, $\rho(\theta)$ and $\rho(\pi-\theta)$. This is depicted in Fig.~\ref{fig1}(a), where light waves are locally reflected back and forth between two nanoscatterer planes within the material. For another wave propagating at an angle of $\pi+\theta$, as depicted in Fig.~\ref{fig1}(b), reciprocity requires that $\tau(\theta) = \tau(\pi+\theta)$ and $\rho(\theta) = \rho(-\theta)$. For a general array of lossy nanoscatterers, the asymmetries $\tau(\theta) \neq \tau(\pi-\theta)$ and $\rho(\theta) \neq \rho(\pi-\theta) = \rho(\pi+\theta)$ can hold. However, for uniaxial nanomaterials considered in this work, the symmetry requires that $\tau(\theta) = \tau(\pi-\theta)$. For bifacial materials, the inequality $\rho(\theta) \neq \rho(\pi+\theta)$ can be the case, which allows the waves propagating in the $\theta$ and $\pi+\theta$ directions to behave differently.

Each unit-cell thick layer in a nanomaterial is in our analysis characterized by the following propagation-direction dependent coefficients
\begin{equation}\label{eqf}
f(\theta) = \tau(\theta) \exp(\mathrm{i} k_{z\mathrm{s}}\Lambda_z),
\end{equation}
\begin{equation}\label{eqg}
g(\theta) = \rho(\theta) \exp(\mathrm{i} k_{z\mathrm{s}}\Lambda_z),
\end{equation}
where $k_{z\mathrm{s}} = k_0 n_\mathrm{s} \cos\theta$ is the z component of the wave vector in the surrounding medium; $k_0$ is the wave number in vacuum. We remind that $\tau$ and $\rho$ depend on the polarization. The first coefficient, $f$, describes the transmission through a single layer of thickness $\Lambda_z$, whereas $g$ describes the reflection from this layer. Between two neighboring layers $j$ and $j+1$, we denote the transverse component of the electric field for the forward and backward propagating waves by $U_j$ and $U_j^{'}$, respectively. These fields are recursively related to each other as\cite{Interfero}
\begin{eqnarray}\label{rel1}
U_{j} &=& f(\theta) U_{j-1}+g(\pi-\theta) U_{j}^{'},\\\label{rel2}
U_j^{'} &=& g(\theta) U_j+f(\theta) U_{j+1}^{'}.
\end{eqnarray}
Using these equations, one can derive
\begin{equation}\label{rec1}
U_{j+1}+U_{j-1}-a U_j = 0,
\end{equation}
where
\begin{equation}\label{eqa}
a = f(\theta) + f(\theta)^{-1}[1-g(\theta)g(\pi-\theta)].
\end{equation}

Next we derive the effective material parameters that characterize the three-dimensional nanoscatterer array as a homogeneous, but spatially dispersive nanomaterial. The first such parameter is the refractive index $n$. In the homogenized material, the transverse electric field of a propagating plane wave must satisfy
\begin{equation}\label{Bloch}
U_{j+1} = U_{j}\exp(\mathrm{i} k_{z}\Lambda_z),
\end{equation}
where $k_{z}$ is the z component of the effective wave vector. Combining Eqs.~(\ref{rec1}) and (\ref{Bloch}), we obtain
\begin{equation}\label{kzL}
k_{z}\Lambda_z = \pm \arccos(a/2)+2\pi m,
\end{equation}
with $m \in \mathbb{Z}$. For nanomaterials which are not bifacial, Eq.~(\ref{kzL}) can be shown to be in perfect agreement with Eq.~(6) of Ref.~\onlinecite{Menzel08}.

The refractive index is obtained as
\begin{equation}\label{neff}
n(\theta) = \pm k_0^{-1}[(k_0 n_\mathrm{s} \sin\theta)^2+k_{z}^2]^{1/2},
\end{equation}
where the continuity of the tangential component of the wave vector is taken into account. This is one of the key results of this work. The choice of order $m$ and signs in Eqs.~(\ref{kzL}) and (\ref{neff}), to obtain physical solutions, are performed as described in Ref.~\onlinecite{Menzel08}, i.e., using the continuity of the $k_{z}$ spectrum and requiring $k_{z}$ and $n$ to have positive imaginary parts. We notice that, since $a$ is symmetric with respect to the interchange between $\theta$ and $\pi-\theta$, \emph{the effective refractive index is the same for any two counter-propagating waves}. This finding can be understood as a consequence of reciprocity for the waves propagating inside the material. As we show next, the same symmetry does not hold for the effective wave impedance.

We derive the wave impedance of the nanomaterial by averaging the electromagnetic fields over a single period of the material in the $z$ direction. Irrespective of the field polarization, we use Eqs.~(\ref{rel1}) and (\ref{Bloch}) to express the component of the electric field that is transverse to the $z$ axis between two neighboring crystal planes as
\begin{eqnarray}\nonumber
E_\perp(z) &=& U_j\exp(\mathrm{i}k_{z\mathrm{s}}z) + U_j^{'}\exp(-\mathrm{i}k_{z\mathrm{s}}z) \\ &=& U_j\{\exp(\mathrm{i}k_{z\mathrm{s}}z)+g(\pi-\theta)^{-1}[1 \nonumber\\
&&-f(\theta)\exp(-\mathrm{i} k_{z}\Lambda_z)]\exp(-\mathrm{i}k_{z\mathrm{s}}z)\}.\label{Etra}
\end{eqnarray}
Since $\mathbf{E}$, $\mathbf{H}$ and $\mathbf{k}_\mathrm{s}$ form a right-handed triad in the host medium, the transverse component of the magnetic field is
\begin{eqnarray}\nonumber
H_\perp(z) &=& \xi\frac{U_j\exp(\mathrm{i}k_{z\mathrm{s}}z)-U_j^{'}\exp(-\mathrm{i}k_{z\mathrm{s}}z)}{\eta_\mathrm{s}}\\ &=& \frac{\xi U_j}{\eta_\mathrm{s}}\{\exp(\mathrm{i}k_{z\mathrm{s}}z)-g(\pi-\theta)^{-1}[1\nonumber\\
&&-f(\theta)\exp(-\mathrm{i} k_{z}\Lambda_z)]\exp(-\mathrm{i}k_{z\mathrm{s}}z)\},\label{Htra}
\end{eqnarray}
where $\eta_\mathrm{s}$ denotes the wave impedance of the host medium and
\begin{equation}
\xi = \begin{cases}
\cos\theta & \text{for TE polarization},\\
1/\cos\theta & \text{for TM polarization}.
\end{cases}
\end{equation}
Notice that we define the reflection coefficient for the TM-polarization such that it has zero phase when the transverse components of the reflected and incident electric fields are in phase. By integrating Eqs.~(\ref{Etra}) and (\ref{Htra}) with respect to $z$ over the interval from $-\Lambda_z/2$ to $\Lambda_z/2$ we can calculate the averaged transverse electric and magnetic fields in the unit cell. Using the fact that $\langle\exp(\mathrm{i}k_{z\mathrm{s}}z)\rangle = \langle\exp(-\mathrm{i}k_{z\mathrm{s}}z)\rangle$, where $\langle\rangle$ denotes the mentioned averaging, we find the ratio between the averaged fields to be
\begin{equation}\label{Ztra}
\frac{\langle E_\perp(z)\rangle}{\langle H_\perp(z)\rangle} = \frac{\eta_\mathrm{s}}{\xi} \frac{g(\pi-\theta)+[1-f(\theta)\exp(-\mathrm{i} k_{z}\Lambda_z)]}{g(\pi-\theta)-[1-f(\theta)\exp(-\mathrm{i} k_{z}\Lambda_z)]}.
\end{equation}
In the homogenized material, the wave (on average) propagates at an angle $\theta_{\mathrm{eff}} \neq \theta$. This angle, as determined by the Snell law $n(\theta)\sin\theta_{\mathrm{eff}} = n_\mathrm{s}\sin\theta$, is obtained from
\begin{equation}\label{thetaeff}
\cos\theta_{\mathrm{eff}} = \pm\Big[1-\frac{n_\mathrm{s}^2}{n(\theta)^2}\sin^2\theta\Big]^{1/2}.
\end{equation}
For the TE polarization, the total magnetic field is found from the transverse magnetic field by dividing it with $\cos\theta_{\mathrm{eff}}$. The total electric field is transverse for this polarization. Using Eq.~(\ref{Ztra}), the effective wave impedance for the TE polarization becomes
\begin{equation}\label{ZTE}
\eta^{\mathrm{TE}}(\theta) = \eta_\mathrm{s}\frac{\cos\theta_{\mathrm{eff}}}{\cos\theta} \frac{g(\pi-\theta)+[1-f(\theta)\exp(-\mathrm{i} k_{z}\Lambda_z)]}{g(\pi-\theta)-[1-f(\theta)\exp(-\mathrm{i} k_{z}\Lambda_z)]}.
\end{equation}
Likewise, for the TM polarization the transverse electric field appearing in the numerator of Eq.~(\ref{Ztra}) must be divided with $\cos\theta_{\mathrm{eff}}$ and, thereby, we obtain
\begin{equation}\label{ZTM}
\eta^{\mathrm{TM}}(\theta) = \eta_\mathrm{s}\frac{\cos\theta}{\cos\theta_{\mathrm{eff}}} \frac{g(\pi-\theta)+[1-f(\theta)\exp(-\mathrm{i}k_{z}\Lambda_z)]}{g(\pi-\theta)-[1-f(\theta)\exp(-\mathrm{i} k_{z}\Lambda_z)]}.
\end{equation}
Notice that the quantities $f$, $g$ and $k_{z}$ in Eqs.~(\ref{ZTE}) and (\ref{ZTM}) have different values for TE and TM polarizations. Obviously, \emph{the impedances obtained for two counter-propagating waves are different in bifacial nanomaterials} [since $\eta(\theta) \neq \eta(\pi+\theta)$, when $g(\pi-\theta) \neq g(-\theta)$].

Equations~(\ref{neff}), (\ref{ZTE}) and (\ref{ZTM}) fully characterize the material. We have verified that for nanomaterials which are not optically bifacial, these equations yield exactly the same results as the retrieval procedure of Ref.~\onlinecite{Menzel08}. The derivation in Ref.~\onlinecite{Menzel08}, however, considers a plane-wave transmission through a nanomaterial slab, whereas the derivation presented here considers the plane waves propagating inside a bulk nanomaterial of an arbitrary shape.

In summary, the material parameters for a given bifacial nanomaterial design can be calculated as follows: The first step is to numerically or otherwise obtain the spectra of $f$ and $g$, defined in Eqs.~(\ref{eqf}) and (\ref{eqg}), for a single two-dimensional layer of nanoscatterers in the host medium. Next, one calculates the spectra of $a$ using Eq.~(\ref{eqa}). Thereafter a continuous spectrum of $k_z$ is calculated from Eq.~(\ref{kzL}). The spectrum of the refractive index follows from Eq.~(\ref{neff}). With the refractive index at hand, the effective propagation angle $\theta_{\mathrm{eff}}$ in the material is solved from Eq.~(\ref{thetaeff}). Finally, depending on the chosen polarization, one uses either Eq.~(\ref{ZTE}) or Eq.~(\ref{ZTM}) to obtain the spectrum of the wave impedance.

\subsection{Transmission and reflection at an interface between two bifacial nanomaterials}\label{SecInt}

The transmission and reflection of an optical plane wave at a material boundary is usually described by using the standard Fresnel coefficients.\cite{Photonics} However, it has been shown that even for strictly homogeneous noncentrosymmetric crystals that are free of spatial dispersion, a generalized form of these coefficients is required.\cite{Graham1996} In the case of optically bifacial nanomaterials, the Fresnel coefficients must be modified as well, to take into account both the asymmetry of the unit cells and the resulting spatial dispersion.

\begin{figure}
\includegraphics[scale=0.8]{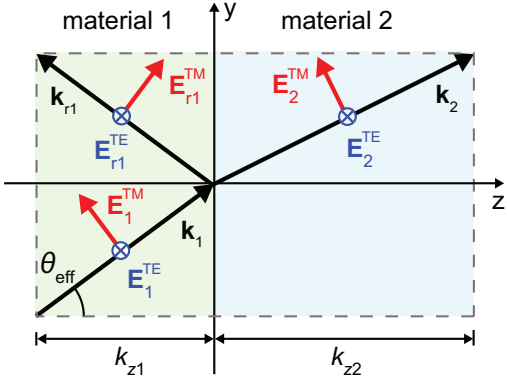}
\centering
\caption{Illustration of a plane wave incident on a boundary between two bifacial materials. In material 1, the incident and reflected waves see the same refractive index $n$, but different wave impedances that are, respectively, denoted by $\eta^\mathrm{R}_1$ and $\eta^\mathrm{L}_1$.\label{figF}}
\end{figure}

Consider a boundary at $z = 0$ between two bifacial materials with material parameters $n_j$ and $\eta_j$, where the index $j \in \{1,2\}$ refers to the material in question. A wave with a wave vector $\mathbf{k}_1 = k_y\hat{\mathbf{y}}+k_{z1}\hat{\mathbf{z}}$ is incident from material 1 onto material 2, in which the wave propagates with the wave vector $\mathbf{k}_2 = k_y\hat{\mathbf{y}}+k_{z2}\hat{\mathbf{z}}$ as depicted in Fig.~\ref{figF}. The magnitude of the wave vector is $k_j = k_0n_j$. For simplicity, we define $\eta^\mathrm{R}_j = \eta_j(\theta)$ and $\eta^\mathrm{L}_j = \eta_j(\pi-\theta)$ to denote the impedances for the waves propagating to the right and left, respectively, in Fig.~\ref{figF}. The vector complex amplitudes of the electric and magnetic fields for the incident ($j = 1$) and transmitted ($j = 2$) waves can be written as
\begin{eqnarray}
\mathbf{E}_j(\mathbf{r}) &=& \big[E_j^{\mathrm{TE}}\hat{\mathbf{x}} \!+\! E_j^{\mathrm{TM}}(\hat{\mathbf{y}}\frac{k_{zj}}{k_j}\!-\!\hat{\mathbf{z}}\frac{k_y}{k_j})\big]\exp(\mathrm{i}\mathbf{k}_j\!\cdot\!\mathbf{r}),\\
\mathbf{H}_j(\mathbf{r}) &=& \big[E_j^{\mathrm{TE}}(\hat{\mathbf{y}}\frac{k_{zj}}{k_j}\!-\!\hat{\mathbf{z}}\frac{k_y}{k_j})\!-\!E_j^{\mathrm{TM}}\hat{\mathbf{x}}\big]\frac{\exp(\mathrm{i}\mathbf{k}_j\!\cdot\!\mathbf{r})}{\eta^\mathrm{R}_j}.
\end{eqnarray}
The wave vector of the reflected wave is $\mathbf{k}_{\mathrm{r1}} = k_y\hat{\mathbf{y}}-k_{z1}\hat{\mathbf{z}}$. The reflected wave is therefore of the form
\begin{eqnarray}
\!\mathbf{E}_{\mathrm{r1}}(\mathbf{r}) &=& \big[E_{\mathrm{r1}}^{\mathrm{TE}}\hat{\mathbf{x}} \!+\!E_{\mathrm{r1}}^{\mathrm{TM}}(\hat{\mathbf{y}}\frac{k_{z1}}{k_1}\!+\!\hat{\mathbf{z}}\frac{k_y}{k_1})\big]\!\exp(\mathrm{i}\mathbf{k}_{\mathrm{r1}}\!\cdot\!\mathbf{r}),\\
\!\mathbf{H}_{\mathrm{r1}}(\mathbf{r}) &=& \big[E_{\mathrm{r1}}^{\mathrm{TM}}\hat{\mathbf{x}} \!-\!E_{\mathrm{r1}}^{\mathrm{TE}}(\hat{\mathbf{y}}\frac{k_{z1}}{k_1}\!+\!\hat{\mathbf{z}}\frac{k_y}{k_1}) \big]\!\frac{\exp(\mathrm{i}\mathbf{k}_{\mathrm{r1}}\!\cdot\!\mathbf{r})}{\eta^\mathrm{L}_1}.
\end{eqnarray}

Applying the electromagnetic boundary conditions that require $\hat{\mathbf{z}}\times(\mathbf{E}_1+\mathbf{E}_{\mathrm{r1}}) = \hat{\mathbf{z}}\times\mathbf{E}_2$ and $\hat{\mathbf{z}}\times(\mathbf{H}_1+\mathbf{H}_{\mathrm{r1}}) = \hat{\mathbf{z}}\times\mathbf{H}_2$ at $z = 0$, we derive the \emph{generalized Fresnel transmission and reflection coefficients} for the TE and TM polarizations as
\begin{eqnarray}\label{Fresnel1}
\tau^{\mathrm{TE}}_{12} = \frac{k_{z1}/(k_1\eta^\mathrm{R}_1)+k_{z1}/(k_1\eta^\mathrm{L}_1)}{k_{z1}/(k_1\eta^\mathrm{L}_1)+k_{z2}/(k_2\eta^\mathrm{R}_2)},\\\label{Fresnel2}
\tau^{\mathrm{TM}}_{12} = \frac{k_{z1}/(k_1\eta^\mathrm{R}_1)+k_{z1}/(k_1\eta^\mathrm{L}_1)}{k_{z1}/(k_1\eta^\mathrm{R}_2)+k_{z2}/(k_2\eta^\mathrm{L}_1)},\\\label{Fresnel3}
\rho^{\mathrm{TE}}_{12} = \frac{k_{z1}/(k_1\eta^\mathrm{R}_1)-k_{z2}/(k_2\eta^\mathrm{R}_2)}{k_{z1}/(k_1\eta^\mathrm{L}_1)+k_{z2}/(k_2\eta^\mathrm{R}_2)},\\
\rho^{\mathrm{TM}}_{12} = \frac{k_{z2}/(k_2\eta^\mathrm{R}_1)-k_{z1}/(k_1\eta^\mathrm{R}_2)}{k_{z1}/(k_1\eta^\mathrm{R}_2)+k_{z2}/(k_2\eta^\mathrm{L}_1)},\label{Fresnel4}
\end{eqnarray}
where the subindex 12 indicates that the field is incident from material 1 onto material 2. As shown in the next section, the bifacial behavior is tightly connected to the excitation of higher-order multipoles. It has previously been suggested that special electromagnetic boundary conditions are required to deal with materials in which higher order multipoles can be excited.\cite{Graham2000} Here, however, the ordinary boundary conditions are applied, since we allow the material parameters to depend on the propagation direction of the plane wave for which they are calculated. This provides a remarkable simplicity for the resulting Fresnel coefficients.

Using Eqs.~(\ref{Fresnel1})-(\ref{Fresnel4}) one can, for instance, calculate the transmission and reflection coefficients of an optically bifacial nanomaterial slab by taking into account multiple reflections. For a wave being incident from a semi-infinite material 1 onto a bifacial slab made of a material 2 of thickness $D$ and transmitted into a semi-infinite material 3, these coefficients are
\begin{eqnarray}\label{eqt}
t &=& \exp(\mathrm{i}k_{z\mathrm{2}}D)\frac{\tau_{12}\tau_{23}}{1-\rho_{21}\rho_{23}\exp(2\mathrm{i}k_{z2}D)},\\\label{eqr}
r &=& \rho_{12} + \frac{\tau_{12}\exp(2\mathrm{i}k_{z2}D)\rho_{23}\tau_{21}}{1-\rho_{21}\rho_{23}\exp(2\mathrm{i}k_{z2}D)}.
\end{eqnarray}
For any bifacial nanomaterial slab, the reflection coefficient $r$ for a wave incident from material 1 differs from that obtained for a wave incident from material 3 even when the materials 1 and 3 are the same, which can be seen by interchanging $\eta^\mathrm{R}_2$ and $\eta^\mathrm{L}_2$ in Eqs.~(\ref{Fresnel1})-(\ref{eqr}). The transmission coefficient, however, is invariant with respect to this interchange, which ensures that the nanomaterial is reciprocal. Equations~(\ref{Fresnel1})-(\ref{eqr}) enable one to verify the effective material parameters calculated by using Eqs.~(\ref{neff}), (\ref{ZTE}) and (\ref{ZTM}). At the end of Sec.~\ref{SecNum}, this verification is performed for a particular bifacial nanomaterial.

\subsection{Electromagnetic multipoles}\label{SecMulti}

In this subsection, we show that for a nanomaterial to be optically bifacial, it is necessary that higher-order electromagnetic multipoles are excited in the nanoscatterers. The transmission and reflection coefficients of a two-dimensional array of nanoscatterers are related to the multipole excitations in each nanoscatterer. Here, for simplicity, we derive these coefficients for normal-incidence illumination of a nanoscatterer array in which both electric dipole and \emph{current} quadrupole\cite{Multipole} moments are excited. Considering an $x$-polarized incident wave, we write the complex amplitude for the $x$ component of the excited electric current density  as
\begin{equation}\label{expJ}
J_x(\mathbf{r}) = -\mathrm{i}\omega\sum_{u,v}\Big(p_x - Q_{xz}\frac{d}{dz}\Big)\delta(\mathbf{r}-u\Lambda\hat{\mathbf{x}}-v\Lambda\hat{\mathbf{y}}),
\end{equation}
where $u$ and $v$ are integers that refer to a certain equivalent point-scatterer in a square lattice of period $\Lambda$ in the $z = 0$ plane. In Eq.~(\ref{expJ}), $p_x$ is the $x$ component of the excited dipole moment and $Q_{xz}$ is the $xz$ element of the current quadrupole dyadic.\cite{Multipole} The contribution from the $xx$ element of the quadrupole dyadic is neglected, as it does not radiate in the $\pm \hat{\mathbf{z}}$ direction. For $\Lambda < \lambda$, no diffraction orders can appear, and Eq.~(\ref{expJ}) can be averaged in the $xy$ plane to obtain
\begin{equation}
\langle J_x(z)\rangle = -\frac{\mathrm{i}\omega}{\Lambda^2}\Big(p_x - Q_{xz}\frac{d}{dz}\Big)\delta(z).
\end{equation}
When replacing $J_x$ with $\langle J_x\rangle$, the evanescent near-field of the array is lost, whereas the far-field that determines the transmission and reflection of the array remains. In the Lorenz gauge, the $x$ component of the vector potential $A_x$ satisfies the wave equation
\begin{equation}\label{wavA}
\Big(\frac{d^2}{dz^2}+k_\mathrm{s}^2\Big)A_x(z) = -\mu_\mathrm{s} \langle J_x(z)\rangle,
\end{equation}
where $k_\mathrm{s}$ and $\mu_\mathrm{s}$ are the wave number and magnetic permeability, respectively, of the surrounding medium. Using the one-dimensional Green's function,\cite{DudleyBook} one can solve Eq.~(\ref{wavA}) for the vector potential and obtain
\begin{equation}\label{Ax}
A_x(z) = \frac{k_\mathrm{s}}{2\omega\epsilon_\mathrm{s}\Lambda^2}\big[p_x-\mathrm{i}k_\mathrm{s}Q_{xz}\mathrm{sign}(z)\big]\exp(\mathrm{i}k_\mathrm{s}|z|).
\end{equation}
The electric far-field scattered by the array, $\mathbf{E}_\mathrm{sca}$, is related to the vector potential through\cite{Multipole}
\begin{equation}\label{Esca}
\mathbf{E}_\mathrm{sca}(\mathbf{r}) = \mathrm{i}\omega\big[\mathbf{A}(\mathbf{r})+\frac{1}{k_\mathrm{s}^2}\nabla\nabla\cdot\mathbf{A}(\mathbf{r})\big].
\end{equation}
The reflection coefficient $\rho$ for the array is obtained from Eqs.~(\ref{Ax}) and (\ref{Esca}) by dividing the complex amplitude $E_\mathrm{sca}$ with the electric field amplitude $E_0$ of the incident wave at $z \to 0^-$. The result is
\begin{equation}\label{rhoMultipoles}
\rho = \frac{\mathrm{i}k_\mathrm{s}}{2\epsilon_\mathrm{s}\Lambda^2}\big[\alpha_x+\mathrm{i}k_\mathrm{s}\beta_{xz}\big].
\end{equation}
The coefficients $\alpha_x = p_x/E_0$ and $\beta_{xz} = Q_{xz}/E_0$ are the electric dipole and current quadrupole \emph{polarizability components} of the scatterers (see also Ref.~\onlinecite{MD2}). For the transmission coefficient $\tau$, the forward scattered field is superposed to the incident field at $z > 0$. Dividing this superposition field with the incident field, we obtain
\begin{equation}\label{tauMultipoles}
\tau = 1 + \frac{\mathrm{i}k_\mathrm{s}}{2\epsilon_\mathrm{s}\Lambda^2}\big[\alpha_x-\mathrm{i}k_\mathrm{s}\beta_{xz}\big].
\end{equation}
Equations (\ref{rhoMultipoles}) and (\ref{tauMultipoles}) give the normal-incidence transmission and reflection coefficients for an arbitrary two-dimensional array of nanoscatterers, provided that the array period is subwavelength-sized. Octupoles and other higher-order multipoles, which may be excited in particular nanoscatterer designs,\cite{Curto13,Kaelberer10122010} can be included by simply adding more terms in the expansion of $J_x$. These multipoles, however, are often negligible for subwavelength-sized nanoparticles.

Important conclusions can now be made from Eqs.~(\ref{rhoMultipoles}) and (\ref{tauMultipoles}) regarding the effect of multipoles on \emph{optically bifacial} and \emph{non-reflective} nanoparticle arrays. The first conclusion is that complete suppression of reflection requires at least the excitation of current quadrupoles in the nanoparticles. We recall that the current quadrupoles include both traditional electric quadrupoles and magnetic dipoles, which are electromagnetic multipoles of the same order.\cite{Multipole} By designing the scatterers such that $p_x = -\mathrm{i}k_\mathrm{s}Q_{xz}$, the reflection can be brought to zero. This is precisely what happens in so-called metamaterial perfect absorbers.\cite{Zeng13} In general, both the transmission and absorption are non-zero. Another conclusion is that the excitation of quadrupoles or other higher-order multipoles are \emph{necessary} for bifacial behavior (for the reflection coefficient to depend on the side of illumination). This is seen by noting that, due to reciprocity, $\tau$ in Eq.~(\ref{tauMultipoles}) must be the same for both illumination sides. Therefore, if $Q_{xz} = 0$, also $p_x$ must be the same for both illumination sides, and so must be the reflection coefficient $\rho$, too.

\section{Numerical studies}\label{SecNum}

\begin{figure}
\includegraphics[scale=0.8]{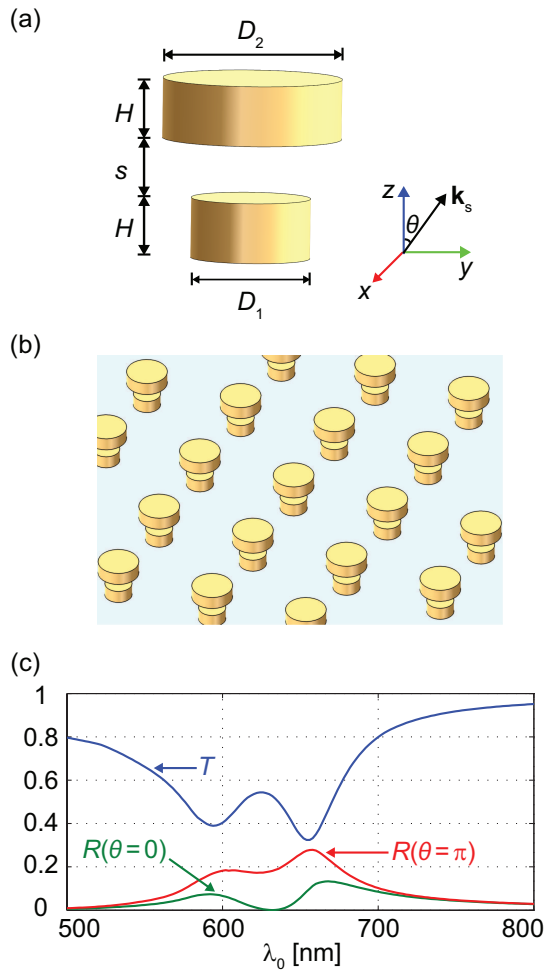}
\centering
\caption{(a) Illustration of a gold disc nanodimer used as an artificial atom in each unit cell of the considered bifacial nanomaterial. (b) Illustration of a two-dimensional nanodimer array spanning a plane within the nanomaterial. (c) Normal-incidence intensity transmission $T$ and reflection $R$ of the array as functions of the wavelength $\lambda_0$ in vacuum. The angles of 0 and $\pi$ correspond to illumination from the side of the smaller and larger discs, respectively.\label{figY}}
\end{figure}

In this section, we use the introduced theory to analyze a particular example of an optically bifacial nanomaterial constructed of gold nanodimers.\cite{MD1,MD2} These nanodimers consist of two axis-aligned discs as depicted in Fig.~\ref{figY}(a). The thickness of both discs is chosen to be $H = 20$~nm and the surface-to-surface separation between the discs is $s = 20$~nm. The diameters of the discs are chosen as $D_1 = 40$~nm and $D_2 = 60$~nm, such that the resonances of the discs occur at different frequencies. In the material, the nanodimers are arranged in a cubic lattice with a period of $\Lambda = 150$~nm and placed in a surrounding dielectric of refractive index $n_\mathrm{s} = 1.5$, corresponding to that of glass. The nanodimer dimensions and the lattice geometry are chosen such that the material is impedance-matched to the surrounding medium when illuminated from the side of the smaller disc.

We start by calculating the complex transmission and reflection coefficients of a \emph{single} two-dimensional array of the nanodimers in glass [see Fig.~\ref{figY}(b)] using the computer software COMSOL Multiphysics. The spectrum of the relative electric permittivity of gold is taken from Ref.~\onlinecite{Johnson1972}. The intensity transmission and reflection spectra of this array at normal incidence are depicted in Fig.~\ref{figY}(c). At a vacuum wavelength of $\lambda_0 = 632$~nm the reflection from the smaller-disc side ($\theta = 0$) is strongly suppressed, whereas the reflection from the other side ($\theta = \pi$) is considerable.

\begin{figure}
\includegraphics[scale=0.8]{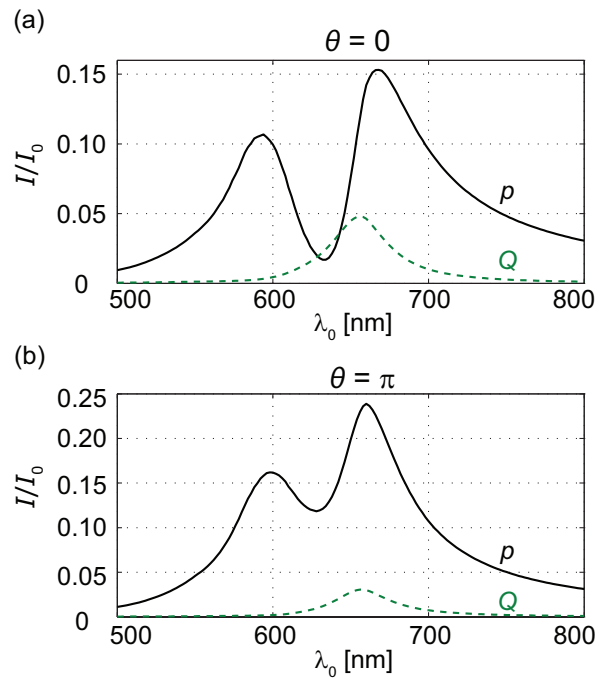}
\centering
\caption{Normalized intensity, $I/I_0$, of the plane wave radiated in the backward direction by the electric dipoles (black solid lines) and current quadrupoles (green dotted lines) excited in the nanodimer array of Fig.~\ref{figY}(b). $I_0$ is the intensity of the incident wave that propagates in the direction of (a) $\theta = 0$ and (b) $\theta = \pi$. The actual reflection coefficient of the array is determined by the interference between the waves created by these two multipoles.\label{figM}}
\end{figure}

Using the theory presented in section~\ref{SecMulti}, we can fully explain this bifacial behavior of the nanodimer array in terms of the electromagnetic multipole excitations. The multipole excitations in the array are extracted from the electric field distribution inside the scatterers as described in Ref.~\onlinecite{Multipole}. The electric dipole and current quadrupole moments are obtained from the multipole coefficients as
\begin{eqnarray}
p_x &=& \frac{6\pi\mathrm{i}\epsilon_\mathrm{s}E_0}{k_\mathrm{s}^3}\Big[\frac{a_{\mathrm{E}}(1,-1)-a_{\mathrm{E}}(1,1)}{2}\Big],\\
Q_{xz} &=& \frac{\pi\epsilon_\mathrm{s}E_0}{k_\mathrm{s}^4}\big[3a_{\mathrm{M}}(1,-1)+3a_{\mathrm{M}}(1,1)\nonumber\\&&
-5a_{\mathrm{E}}(2,-1)+5a_{\mathrm{E}}(2,1)\big],
\end{eqnarray}
where $a_{\mathrm{E}}$ and $a_{\mathrm{M}}$ are the electric and magnetic multipole coefficients, respectively, described in detail in Ref.~\onlinecite{Multipole}. In Fig.~\ref{figM}, we show the normalized intensity of the field radiated in the backward direction separately by the dipoles and current quadrupoles, as obtained by calculating the squared moduli of the individual terms in Eq.~(\ref{rhoMultipoles}). The actual reflection shown in Fig.~\ref{figY}(c) is the result of interference between the two multipole waves as given by Eq.~(\ref{rhoMultipoles}). The magnitudes of multipoles of higher orders than the current quadrupole are negligibly small. From Fig.~\ref{figM}(a), we see that in the region of $\lambda_0 = 632$~nm, the dipole and quadrupole waves have equal amplitudes. These two waves oscillate out of phase and interfere destructively, providing the suppressed reflection in Fig.~\ref{figY}(c). In contrast, for the opposite propagation direction, $\theta = \pi$, the dipole contribution is much larger than the quadrupole one over the whole spectral range considered [see Fig.~\ref{figM}(b)]. Consequently, the reflection is significant everywhere in this wavelength range.

In order to calculate the effective electromagnetic material parameters for the considered three-dimensional nanomaterial, we use Eqs.~(\ref{neff}), (\ref{ZTE}) and (\ref{ZTM}) with the complex transmission and reflection coefficients that are numerically obtained for the two-dimensional array of the dimers. The effective refractive index and impedance are expected to depend on both the angle $\theta$ and the polarization direction of the light. Consequently, we consider both TE- and TM-polarized waves and a large enough set of propagation angles $\theta$ in the host medium, choosing the values of 0, $\pi/6$, $\pi/3$, $2\pi/3$, $5\pi/6$, and $\pi$. As explained in Sec.~\ref{SecMP}, it is sufficient to evaluate the refractive index only for $\theta \leq \pi/2$, because of the symmetry $n(\pi-\theta) = n(\theta)$. Before discussing our results, we emphasize that we have verified the obtained material parameters by doing rigorous numerical calculations for a five-layer thick nanomaterial slab and by comparing those calculations with the results obtained by applying Eqs.~(\ref{Fresnel1})-(\ref{eqr}) with the evaluated material parameters. The agreement is perfect, as will be shown later on in this section.

\begin{figure}
\includegraphics[scale=0.8]{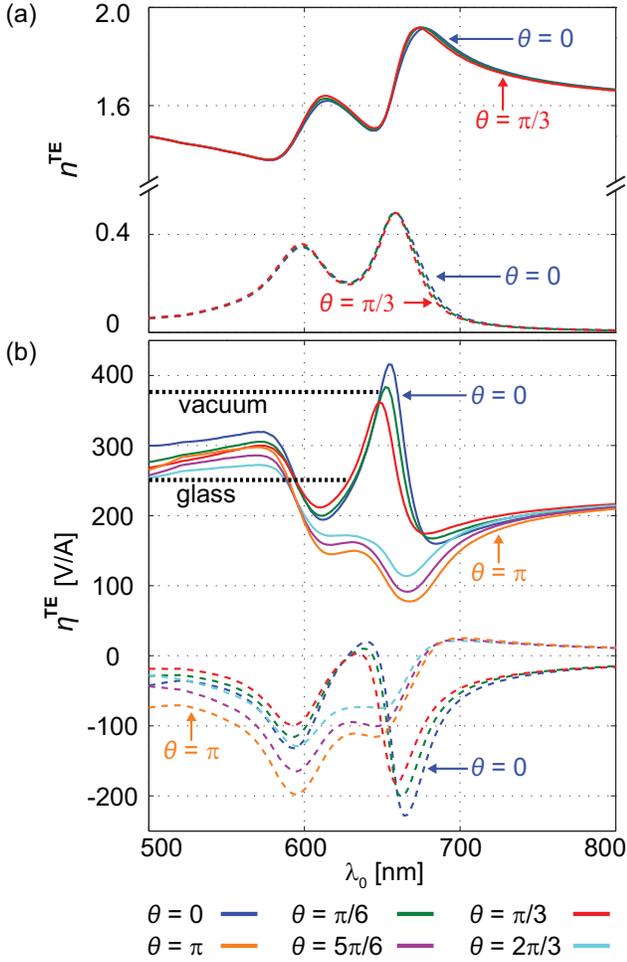}
\centering
\caption{(a) Refractive index and (b) wave impedance spectra of the nanodimer nanomaterial for TE polarized light. The real parts (solid lines) and imaginary parts (dashed lines) of the quantities are shown separately. Different propagation angles $\theta$ in the host dielectric are marked with different colors. The wave impedances of vacuum and glass (refractive index 1.5) are shown in (b) by the horizontal black dashed lines.\label{figTE}}
\end{figure}

For the TE polarization, with the electric field perpendicular to the dimer axis, the calculated spectra of the refractive index and the wave impedance are shown in Figs.~\ref{figTE}(a) and \ref{figTE}(b), respectively. The real parts of these complex quantities are shown by solid lines, while the imaginary parts are shown by dashed lines. It can be seen that, for this polarization, the refractive index is nearly independent of the light propagation direction. This feature can be understood by noting that the modal excitations in the individual discs can only slightly depend on the angle of incidence of the light. In the refractive index spectra, one can clearly distinguish the dipole resonances of the two discs composing the dimers.

\begin{figure}
\includegraphics[scale=0.8]{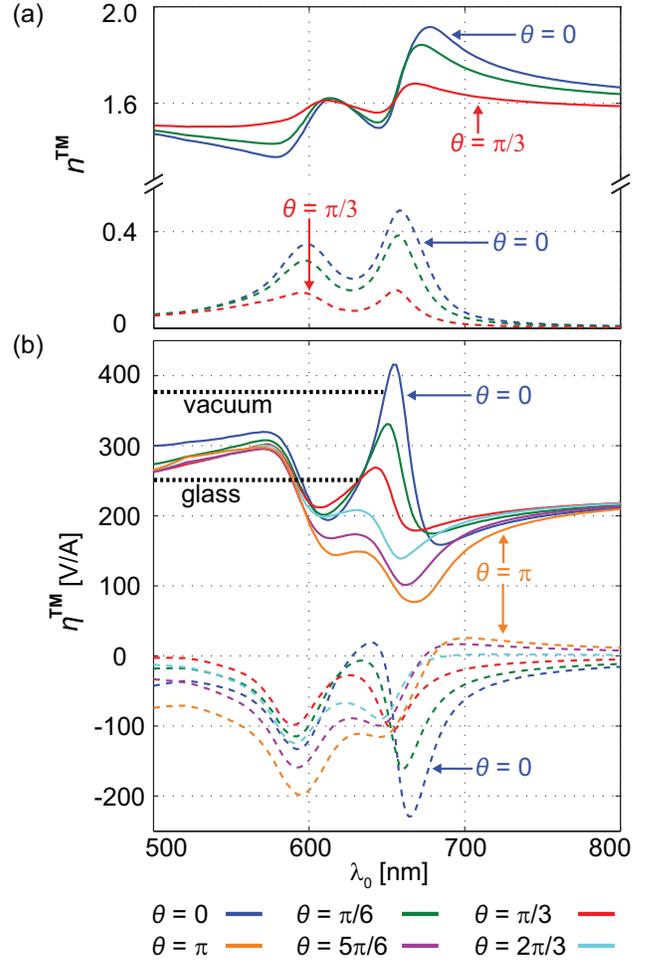}
\centering
\caption{(a) Refractive index and (b) wave impedance spectra of the nanodimer nanomaterial for TM polarized light. The real parts (solid lines) and imaginary parts (dashed lines) of the quantities are shown separately. Different propagation angles $\theta$ in the host dielectric are marked with different colors. The wave impedances of vacuum and glass (refractive index 1.5) are shown in (b) by the horizontal black dashed lines.\label{figTM}}
\end{figure}

In contrast to the refractive index, the wave impedance, as expected, depends on the propagation direction quite significantly [see Fig.~\ref{figTE}(b)]. At a wavelength of about 630~nm, the wave impedance for $\theta \in \{0,\pi/6,\pi/3\}$ is essentially real-valued and quite well matched to that of the surrounding glass, which means that the light reflection at a glass-nanomaterial interface will be suppressed [see Eq.~(\ref{Fresnel3})]. The smaller discs are in this case pointed towards the interface. On the other hand, if the larger discs are those which are closer to the interface, the interface reflects light significantly, as follows from a completely different value of wave impedance for $\theta \in \{\pi,5\pi/6,2\pi/3\}$ at this wavelength. Still one can see that in the range of $\theta \in [-\pi/3,\pi/3]$ (or $\theta \in [2\pi/3, 4\pi/3]$) the material can be considered to be nearly spatially non-dispersive. It is nonetheless quite obvious that the material is optically bifacial.

The calculated spectra for the refractive index and the wave impedance corresponding to TM polarized light are shown in Fig.~\ref{figTM}. For this polarization, the refractive index changes significantly with the propagation angle $\theta$, as shown in Fig.~\ref{figTM}(a). This is a consequence of the fact that the $z$ component of the electric field interacts non-resonantly with the nanodimers. Hence, for angles $\theta$ approaching $\pi/2$, the resonant interaction of the wave with the dimers gradually disappears. For the propagation directions $\theta = 0$ and $\theta = \pi$, the wave impedances are as expected identical to those for the TE polarization. However, for the TM polarization, the wave impedance changes more with increasing $\theta$ than for the TE polarization. At $\lambda_0 \approx 630$~nm, we still have quite good impedance-matching of the material to glass at $\theta \in \{0,\pi/6,\pi/3\}$.

\begin{figure}
\includegraphics[scale=0.8]{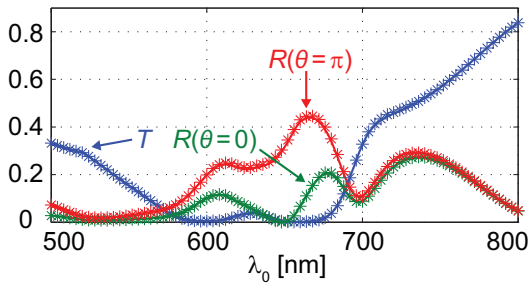}
\centering
\caption{Normal-incidence intensity transmission $T$ and reflection $R$ of a nanomaterial slab composed of 5 layers of nanodimers embedded in glass. The slab is located in vacuum. The lines show the results obtained by using Eqs.~(\ref{Fresnel1})-(\ref{eqr}) with the material parameters obtained from a single two-dimensional array in glass. The stars show the results of direct numerical calculation.\label{figA5}}
\end{figure}

We have verified the validity of the obtained material parameters and the expressions for them in Eqs.~(\ref{neff}), (\ref{ZTE}) and (\ref{ZTM}) by using Eqs.~(\ref{Fresnel1})-(\ref{eqr}) to calculate the transmission and reflection of light by a nanomaterial slab \emph{in vacuum}. The slab has a thickness of 750~nm and contains five nanodimer layers [each layer is as the one in Fig.~\ref{figY}(b)] embedded in glass. The calculated intensity transmission and reflection spectra of this slab are depicted by solid lines in Fig.~\ref{figA5}. For comparison, we use COMSOL to directly calculate the transmission and reflection by the slab. The results of these calculations are presented in Fig.~\ref{figA5} by stars. These results are seen to be in perfect agreement with the results obtained using our effective material parameters. The suppressed reflection at $\lambda_0 \approx 650$~nm originates from the fact that at this wavelength the nanomaterial is approximately impedance-matched to vacuum (see Figs.~\ref{figTE} and \ref{figTM}).

\section{Conclusions}

We have introduced a theoretical model for calculating the electromagnetic characteristics of designed bifacial optical nanomaterials that are composed of noncentrosymmetric nanoparticles with a single symmetry axis. In general, these characteristics depend on the propagation direction of light, and for bifacial nanomaterials they are different for two counter-propagating waves. Our approach enables direct evaluation of the material's refractive index and wave impedance from the transmission and reflection coefficients of a single crystal plane of the constituent nanoscatterers. This makes the required numerical calculations fast. Taking into account the dependence of the material parameters on the light propagation direction, we have derived generalized Fresnel coefficients that describe the light transmission and reflection at the surface of a bifacial nanomaterial. We have found that in order to create an optically bifacial nanomaterial, it is necessary to use nanoscatterers in which light can efficiently excite higher-order electromagnetic multipoles. One remarkable finding is that inside such a nanomaterial, two counter-propagating waves necessarily see equal refractive indices, while they can experience quite different wave impedances.

The practical application of the theory was demonstrated with a particular example of a bifacial nanomaterial composed of metal-disc nanodimers. Using the model, we have designed and characterized a material that is impedance-matched to the surrounding medium (glass and vacuum) for a wave propagating in a certain direction, but not for a counter-propagating wave.

This work opens up the possibility to comprehensively describe and design uniaxial bifacial optical nanomaterials. The asymmetry of the unit cells in such materials can provide the substance with an extraordinary electromagnetic response that is not possible to find in natural materials. For example, a nanomaterial slab can be designed to have a desired spectrum of the refractive index and/or to reflect light only by one of its surfaces. The latter feature can be of interest for imaging and energy-harvesting applications, including ultrathin wide-angle bifacial antireflection coatings and high-efficiency solar cells.

\begin{acknowledgments}
This work was funded by the Academy of Finland (Project No. 134029).
\end{acknowledgments}

%\bibliography{library}

\begin{thebibliography}{34}%
\makeatletter
\providecommand \@ifxundefined [1]{%
 \@ifx{#1\undefined}
}%
\providecommand \@ifnum [1]{%
 \ifnum #1\expandafter \@firstoftwo
 \else \expandafter \@secondoftwo
 \fi
}%
\providecommand \@ifx [1]{%
 \ifx #1\expandafter \@firstoftwo
 \else \expandafter \@secondoftwo
 \fi
}%
\providecommand \natexlab [1]{#1}%
\providecommand \enquote  [1]{``#1''}%
\providecommand \bibnamefont  [1]{#1}%
\providecommand \bibfnamefont [1]{#1}%
\providecommand \citenamefont [1]{#1}%
\providecommand \href@noop [0]{\@secondoftwo}%
\providecommand \href [0]{\begingroup \@sanitize@url \@href}%
\providecommand \@href[1]{\@@startlink{#1}\@@href}%
\providecommand \@@href[1]{\endgroup#1\@@endlink}%
\providecommand \@sanitize@url [0]{\catcode `\\12\catcode `\$12\catcode
  `\&12\catcode `\#12\catcode `\^12\catcode `\_12\catcode `\%12\relax}%
\providecommand \@@startlink[1]{}%
\providecommand \@@endlink[0]{}%
\providecommand \url  [0]{\begingroup\@sanitize@url \@url }%
\providecommand \@url [1]{\endgroup\@href {#1}{\urlprefix }}%
\providecommand \urlprefix  [0]{URL }%
\providecommand \Eprint [0]{\href }%
\providecommand \doibase [0]{http://dx.doi.org/}%
\providecommand \selectlanguage [0]{\@gobble}%
\providecommand \bibinfo  [0]{\@secondoftwo}%
\providecommand \bibfield  [0]{\@secondoftwo}%
\providecommand \translation [1]{[#1]}%
\providecommand \BibitemOpen [0]{}%
\providecommand \bibitemStop [0]{}%
\providecommand \bibitemNoStop [0]{.\EOS\space}%
\providecommand \EOS [0]{\spacefactor3000\relax}%
\providecommand \BibitemShut  [1]{\csname bibitem#1\endcsname}%
\let\auto@bib@innerbib\@empty
%</preamble>
\bibitem [{\citenamefont {Menzel}\ \emph
  {et~al.}(2008{\natexlab{a}})\citenamefont {Menzel}, \citenamefont
  {Rockstuhl}, \citenamefont {Paul}, \citenamefont {Lederer},\ and\
  \citenamefont {Pertsch}}]{Menzel08}%
  \BibitemOpen
  \bibfield  {author} {\bibinfo {author} {\bibfnamefont {C.}~\bibnamefont
  {Menzel}}, \bibinfo {author} {\bibfnamefont {C.}~\bibnamefont {Rockstuhl}},
  \bibinfo {author} {\bibfnamefont {T.}~\bibnamefont {Paul}}, \bibinfo {author}
  {\bibfnamefont {F.}~\bibnamefont {Lederer}}, \ and\ \bibinfo {author}
  {\bibfnamefont {T.}~\bibnamefont {Pertsch}},\ }\href {\doibase
  10.1103/PhysRevB.77.195328} {\bibfield  {journal} {\bibinfo  {journal} {Phys.
  Rev. B}\ }\textbf {\bibinfo {volume} {77}},\ \bibinfo {pages} {195328}
  (\bibinfo {year} {2008}{\natexlab{a}})}\BibitemShut {NoStop}%
\bibitem [{\citenamefont {Smith}\ \emph {et~al.}(2002)\citenamefont {Smith},
  \citenamefont {Schultz}, \citenamefont {Marko\ifmmode~\check{s}\else
  \v{s}\fi{}},\ and\ \citenamefont {Soukoulis}}]{Smith02}%
  \BibitemOpen
  \bibfield  {author} {\bibinfo {author} {\bibfnamefont {D.~R.}\ \bibnamefont
  {Smith}}, \bibinfo {author} {\bibfnamefont {S.}~\bibnamefont {Schultz}},
  \bibinfo {author} {\bibfnamefont {P.}~\bibnamefont
  {Marko\ifmmode~\check{s}\else \v{s}\fi{}}}, \ and\ \bibinfo {author}
  {\bibfnamefont {C.~M.}\ \bibnamefont {Soukoulis}},\ }\href {\doibase
  10.1103/PhysRevB.65.195104} {\bibfield  {journal} {\bibinfo  {journal} {Phys.
  Rev. B}\ }\textbf {\bibinfo {volume} {65}},\ \bibinfo {pages} {195104}
  (\bibinfo {year} {2002})}\BibitemShut {NoStop}%
\bibitem [{\citenamefont {Menzel}\ \emph
  {et~al.}(2008{\natexlab{b}})\citenamefont {Menzel}, \citenamefont
  {Rockstuhl}, \citenamefont {Paul},\ and\ \citenamefont
  {Lederer}}]{Menzel08-2}%
  \BibitemOpen
  \bibfield  {author} {\bibinfo {author} {\bibfnamefont {C.}~\bibnamefont
  {Menzel}}, \bibinfo {author} {\bibfnamefont {C.}~\bibnamefont {Rockstuhl}},
  \bibinfo {author} {\bibfnamefont {T.}~\bibnamefont {Paul}}, \ and\ \bibinfo
  {author} {\bibfnamefont {F.}~\bibnamefont {Lederer}},\ }\href {\doibase
  10.1063/1.3046127} {\bibfield  {journal} {\bibinfo  {journal} {Applied
  Physics Letters}\ }\textbf {\bibinfo {volume} {93}},\ \bibinfo {eid} {233106}
  (\bibinfo {year} {2008}{\natexlab{b}})}\BibitemShut {NoStop}%
\bibitem [{\citenamefont {Pshenay-Severin}\ \emph {et~al.}(2009)\citenamefont
  {Pshenay-Severin}, \citenamefont {H\"{u}bner}, \citenamefont {Menzel},
  \citenamefont {Helgert}, \citenamefont {Chipouline}, \citenamefont
  {Rockstuhl}, \citenamefont {T\"{u}nnermann}, \citenamefont {Lederer},\ and\
  \citenamefont {Pertsch}}]{Pshenay09}%
  \BibitemOpen
  \bibfield  {author} {\bibinfo {author} {\bibfnamefont {E.}~\bibnamefont
  {Pshenay-Severin}}, \bibinfo {author} {\bibfnamefont {U.}~\bibnamefont
  {H\"{u}bner}}, \bibinfo {author} {\bibfnamefont {C.}~\bibnamefont {Menzel}},
  \bibinfo {author} {\bibfnamefont {C.}~\bibnamefont {Helgert}}, \bibinfo
  {author} {\bibfnamefont {A.}~\bibnamefont {Chipouline}}, \bibinfo {author}
  {\bibfnamefont {C.}~\bibnamefont {Rockstuhl}}, \bibinfo {author}
  {\bibfnamefont {A.}~\bibnamefont {T\"{u}nnermann}}, \bibinfo {author}
  {\bibfnamefont {F.}~\bibnamefont {Lederer}}, \ and\ \bibinfo {author}
  {\bibfnamefont {T.}~\bibnamefont {Pertsch}},\ }\href {\doibase
  10.1364/OL.34.001678} {\bibfield  {journal} {\bibinfo  {journal} {Opt.
  Lett.}\ }\textbf {\bibinfo {volume} {34}},\ \bibinfo {pages} {1678} (\bibinfo
  {year} {2009})}\BibitemShut {NoStop}%
\bibitem [{\citenamefont {Choi}\ \emph {et~al.}(2011)\citenamefont {Choi},
  \citenamefont {Lee}, \citenamefont {Kim}, \citenamefont {Kang}, \citenamefont
  {Shin}, \citenamefont {Kwak}, \citenamefont {Kang}, \citenamefont {Lee},
  \citenamefont {Park},\ and\ \citenamefont {Min}}]{Choi11}%
  \BibitemOpen
  \bibfield  {author} {\bibinfo {author} {\bibfnamefont {M.}~\bibnamefont
  {Choi}}, \bibinfo {author} {\bibfnamefont {S.~H.}\ \bibnamefont {Lee}},
  \bibinfo {author} {\bibfnamefont {Y.}~\bibnamefont {Kim}}, \bibinfo {author}
  {\bibfnamefont {S.~B.}\ \bibnamefont {Kang}}, \bibinfo {author}
  {\bibfnamefont {J.}~\bibnamefont {Shin}}, \bibinfo {author} {\bibfnamefont
  {M.~H.}\ \bibnamefont {Kwak}}, \bibinfo {author} {\bibfnamefont {K.-Y.}\
  \bibnamefont {Kang}}, \bibinfo {author} {\bibfnamefont {Y.-H.}\ \bibnamefont
  {Lee}}, \bibinfo {author} {\bibfnamefont {N.}~\bibnamefont {Park}}, \ and\
  \bibinfo {author} {\bibfnamefont {B.}~\bibnamefont {Min}},\ }\href {\doibase
  10.1038/nature09776} {\bibfield  {journal} {\bibinfo  {journal} {Nature}\
  }\textbf {\bibinfo {volume} {470}},\ \bibinfo {pages} {369} (\bibinfo {year}
  {2011})}\BibitemShut {NoStop}%
\bibitem [{\citenamefont {Silveirinha}\ and\ \citenamefont
  {Engheta}(2007)}]{Silveirinha07}%
  \BibitemOpen
  \bibfield  {author} {\bibinfo {author} {\bibfnamefont {M.}~\bibnamefont
  {Silveirinha}}\ and\ \bibinfo {author} {\bibfnamefont {N.}~\bibnamefont
  {Engheta}},\ }\href {\doibase 10.1103/PhysRevB.75.075119} {\bibfield
  {journal} {\bibinfo  {journal} {Phys. Rev. B}\ }\textbf {\bibinfo {volume}
  {75}},\ \bibinfo {pages} {075119} (\bibinfo {year} {2007})}\BibitemShut
  {NoStop}%
\bibitem [{\citenamefont {Yuan}\ \emph {et~al.}(2007)\citenamefont {Yuan},
  \citenamefont {Chettiar}, \citenamefont {Cai}, \citenamefont {Kildishev},
  \citenamefont {Boltasseva}, \citenamefont {Drachev},\ and\ \citenamefont
  {Shalaev}}]{Yuan07}%
  \BibitemOpen
  \bibfield  {author} {\bibinfo {author} {\bibfnamefont {H.-K.}\ \bibnamefont
  {Yuan}}, \bibinfo {author} {\bibfnamefont {U.~K.}\ \bibnamefont {Chettiar}},
  \bibinfo {author} {\bibfnamefont {W.}~\bibnamefont {Cai}}, \bibinfo {author}
  {\bibfnamefont {A.~V.}\ \bibnamefont {Kildishev}}, \bibinfo {author}
  {\bibfnamefont {A.}~\bibnamefont {Boltasseva}}, \bibinfo {author}
  {\bibfnamefont {V.~P.}\ \bibnamefont {Drachev}}, \ and\ \bibinfo {author}
  {\bibfnamefont {V.~M.}\ \bibnamefont {Shalaev}},\ }\href {\doibase
  10.1364/OE.15.001076} {\bibfield  {journal} {\bibinfo  {journal} {Opt.
  Express}\ }\textbf {\bibinfo {volume} {15}},\ \bibinfo {pages} {1076}
  (\bibinfo {year} {2007})}\BibitemShut {NoStop}%
\bibitem [{\citenamefont {Rockstuhl}\ \emph {et~al.}(2009)\citenamefont
  {Rockstuhl}, \citenamefont {Menzel}, \citenamefont {Paul},\ and\
  \citenamefont {Lederer}}]{Rockstuhl09}%
  \BibitemOpen
  \bibfield  {author} {\bibinfo {author} {\bibfnamefont {C.}~\bibnamefont
  {Rockstuhl}}, \bibinfo {author} {\bibfnamefont {C.}~\bibnamefont {Menzel}},
  \bibinfo {author} {\bibfnamefont {T.}~\bibnamefont {Paul}}, \ and\ \bibinfo
  {author} {\bibfnamefont {F.}~\bibnamefont {Lederer}},\ }\href {\doibase
  10.1103/PhysRevB.79.035321} {\bibfield  {journal} {\bibinfo  {journal} {Phys.
  Rev. B}\ }\textbf {\bibinfo {volume} {79}},\ \bibinfo {pages} {035321}
  (\bibinfo {year} {2009})}\BibitemShut {NoStop}%
\bibitem [{\citenamefont {Paul}\ \emph {et~al.}(2009)\citenamefont {Paul},
  \citenamefont {Rockstuhl}, \citenamefont {Menzel},\ and\ \citenamefont
  {Lederer}}]{Paul09}%
  \BibitemOpen
  \bibfield  {author} {\bibinfo {author} {\bibfnamefont {T.}~\bibnamefont
  {Paul}}, \bibinfo {author} {\bibfnamefont {C.}~\bibnamefont {Rockstuhl}},
  \bibinfo {author} {\bibfnamefont {C.}~\bibnamefont {Menzel}}, \ and\ \bibinfo
  {author} {\bibfnamefont {F.}~\bibnamefont {Lederer}},\ }\href {\doibase
  10.1103/PhysRevB.79.115430} {\bibfield  {journal} {\bibinfo  {journal} {Phys.
  Rev. B}\ }\textbf {\bibinfo {volume} {79}},\ \bibinfo {pages} {115430}
  (\bibinfo {year} {2009})}\BibitemShut {NoStop}%
\bibitem [{\citenamefont {Li}\ \emph {et~al.}(2013)\citenamefont {Li},
  \citenamefont {Mutlu},\ and\ \citenamefont {Ozbay}}]{Li13}%
  \BibitemOpen
  \bibfield  {author} {\bibinfo {author} {\bibfnamefont {Z.}~\bibnamefont
  {Li}}, \bibinfo {author} {\bibfnamefont {M.}~\bibnamefont {Mutlu}}, \ and\
  \bibinfo {author} {\bibfnamefont {E.}~\bibnamefont {Ozbay}},\ }\href
  {\doibase 10.1088/2040-8978/15/2/023001} {\bibfield  {journal} {\bibinfo
  {journal} {Journal of Optics}\ }\textbf {\bibinfo {volume} {15}},\ \bibinfo
  {pages} {023001} (\bibinfo {year} {2013})}\BibitemShut {NoStop}%
\bibitem [{\citenamefont {Gompf}\ \emph {et~al.}(2012)\citenamefont {Gompf},
  \citenamefont {Krausz}, \citenamefont {Frank},\ and\ \citenamefont
  {Dressel}}]{Gompf2012}%
  \BibitemOpen
  \bibfield  {author} {\bibinfo {author} {\bibfnamefont {B.}~\bibnamefont
  {Gompf}}, \bibinfo {author} {\bibfnamefont {B.}~\bibnamefont {Krausz}},
  \bibinfo {author} {\bibfnamefont {B.}~\bibnamefont {Frank}}, \ and\ \bibinfo
  {author} {\bibfnamefont {M.}~\bibnamefont {Dressel}},\ }\href {\doibase
  10.1103/PhysRevB.86.075462} {\bibfield  {journal} {\bibinfo  {journal} {Phys.
  Rev. B}\ }\textbf {\bibinfo {volume} {86}},\ \bibinfo {pages} {075462}
  (\bibinfo {year} {2012})}\BibitemShut {NoStop}%
\bibitem [{\citenamefont {Grahn}\ \emph
  {et~al.}(2013{\natexlab{a}})\citenamefont {Grahn}, \citenamefont
  {Shevchenko},\ and\ \citenamefont {Kaivola}}]{MD2}%
  \BibitemOpen
  \bibfield  {author} {\bibinfo {author} {\bibfnamefont {P.}~\bibnamefont
  {Grahn}}, \bibinfo {author} {\bibfnamefont {A.}~\bibnamefont {Shevchenko}}, \
  and\ \bibinfo {author} {\bibfnamefont {M.}~\bibnamefont {Kaivola}},\ }\href
  {\doibase 10.2971/jeos.2013.13009} {\bibfield  {journal} {\bibinfo  {journal}
  {J. Europ. Opt. Soc. Rap. Public.}\ }\textbf {\bibinfo {volume} {8}},\
  \bibinfo {pages} {13009} (\bibinfo {year} {2013}{\natexlab{a}})}\BibitemShut
  {NoStop}%
\bibitem [{\citenamefont {Pakizeh}\ and\ \citenamefont
  {K\"{a}ll}(2009)}]{Pakizeh09}%
  \BibitemOpen
  \bibfield  {author} {\bibinfo {author} {\bibfnamefont {T.}~\bibnamefont
  {Pakizeh}}\ and\ \bibinfo {author} {\bibfnamefont {M.}~\bibnamefont
  {K\"{a}ll}},\ }\href {\doibase 10.1021/nl900786u} {\bibfield  {journal}
  {\bibinfo  {journal} {Nano Letters}\ }\textbf {\bibinfo {volume} {9}},\
  \bibinfo {pages} {2343} (\bibinfo {year} {2009})}\BibitemShut {NoStop}%
\bibitem [{\citenamefont {Pakizeh}\ \emph {et~al.}(2008)\citenamefont
  {Pakizeh}, \citenamefont {Dmitriev}, \citenamefont {Abrishamian},
  \citenamefont {Granpayeh},\ and\ \citenamefont {K\"{a}ll}}]{Pakizeh2008}%
  \BibitemOpen
  \bibfield  {author} {\bibinfo {author} {\bibfnamefont {T.}~\bibnamefont
  {Pakizeh}}, \bibinfo {author} {\bibfnamefont {A.}~\bibnamefont {Dmitriev}},
  \bibinfo {author} {\bibfnamefont {M.~S.}\ \bibnamefont {Abrishamian}},
  \bibinfo {author} {\bibfnamefont {N.}~\bibnamefont {Granpayeh}}, \ and\
  \bibinfo {author} {\bibfnamefont {M.}~\bibnamefont {K\"{a}ll}},\ }\href
  {\doibase 10.1364/JOSAB.25.000659} {\bibfield  {journal} {\bibinfo  {journal}
  {J. Opt. Soc. Am. B}\ }\textbf {\bibinfo {volume} {25}},\ \bibinfo {pages}
  {659} (\bibinfo {year} {2008})}\BibitemShut {NoStop}%
\bibitem [{\citenamefont {Pakizeh}(2012)}]{Pakizeh12}%
  \BibitemOpen
  \bibfield  {author} {\bibinfo {author} {\bibfnamefont {T.}~\bibnamefont
  {Pakizeh}},\ }\href {\doibase 10.1364/JOSAB.29.002446} {\bibfield  {journal}
  {\bibinfo  {journal} {J. Opt. Soc. Am. B}\ }\textbf {\bibinfo {volume}
  {29}},\ \bibinfo {pages} {2446} (\bibinfo {year} {2012})}\BibitemShut
  {NoStop}%
\bibitem [{\citenamefont {Antosiewicz}\ \emph {et~al.}(2012)\citenamefont
  {Antosiewicz}, \citenamefont {Apell}, \citenamefont {Wadell},\ and\
  \citenamefont {Langhammer}}]{Anto12}%
  \BibitemOpen
  \bibfield  {author} {\bibinfo {author} {\bibfnamefont {T.~J.}\ \bibnamefont
  {Antosiewicz}}, \bibinfo {author} {\bibfnamefont {S.~P.}\ \bibnamefont
  {Apell}}, \bibinfo {author} {\bibfnamefont {C.}~\bibnamefont {Wadell}}, \
  and\ \bibinfo {author} {\bibfnamefont {C.}~\bibnamefont {Langhammer}},\
  }\href {\doibase 10.1021/jp306541n} {\bibfield  {journal} {\bibinfo
  {journal} {The Journal of Physical Chemistry C}\ }\textbf {\bibinfo {volume}
  {116}},\ \bibinfo {pages} {20522} (\bibinfo {year} {2012})}\BibitemShut
  {NoStop}%
\bibitem [{\citenamefont {Shegai}\ \emph {et~al.}(2011)\citenamefont {Shegai},
  \citenamefont {Chen}, \citenamefont {Miljkovi\'c}, \citenamefont {Zengin},
  \citenamefont {Johansson},\ and\ \citenamefont {K\"all}}]{Shegai11}%
  \BibitemOpen
  \bibfield  {author} {\bibinfo {author} {\bibfnamefont {T.}~\bibnamefont
  {Shegai}}, \bibinfo {author} {\bibfnamefont {S.}~\bibnamefont {Chen}},
  \bibinfo {author} {\bibfnamefont {V.~D.}\ \bibnamefont {Miljkovi\'c}},
  \bibinfo {author} {\bibfnamefont {G.}~\bibnamefont {Zengin}}, \bibinfo
  {author} {\bibfnamefont {P.}~\bibnamefont {Johansson}}, \ and\ \bibinfo
  {author} {\bibfnamefont {M.}~\bibnamefont {K\"all}},\ }\href {\doibase
  10.1038/ncomms1490} {\bibfield  {journal} {\bibinfo  {journal} {Nature
  Commun.}\ }\textbf {\bibinfo {volume} {2}},\ \bibinfo {pages} {481} (\bibinfo
  {year} {2011})}\BibitemShut {NoStop}%
\bibitem [{\citenamefont {Lavasani}\ and\ \citenamefont
  {Pakizeh}(2012)}]{AlaviLavasani12}%
  \BibitemOpen
  \bibfield  {author} {\bibinfo {author} {\bibfnamefont {S.~H.~A.}\
  \bibnamefont {Lavasani}}\ and\ \bibinfo {author} {\bibfnamefont
  {T.}~\bibnamefont {Pakizeh}},\ }\href {\doibase 10.1364/JOSAB.29.001361}
  {\bibfield  {journal} {\bibinfo  {journal} {J. Opt. Soc. Am. B}\ }\textbf
  {\bibinfo {volume} {29}},\ \bibinfo {pages} {1361} (\bibinfo {year}
  {2012})}\BibitemShut {NoStop}%
\bibitem [{\citenamefont {Liu}\ \emph {et~al.}(2009)\citenamefont {Liu},
  \citenamefont {Langguth}, \citenamefont {Weiss}, \citenamefont {K\"astel},
  \citenamefont {Fleischhauer}, \citenamefont {Pfau},\ and\ \citenamefont
  {Giessen}}]{Liu09}%
  \BibitemOpen
  \bibfield  {author} {\bibinfo {author} {\bibfnamefont {N.}~\bibnamefont
  {Liu}}, \bibinfo {author} {\bibfnamefont {L.}~\bibnamefont {Langguth}},
  \bibinfo {author} {\bibfnamefont {T.}~\bibnamefont {Weiss}}, \bibinfo
  {author} {\bibfnamefont {J.}~\bibnamefont {K\"astel}}, \bibinfo {author}
  {\bibfnamefont {M.}~\bibnamefont {Fleischhauer}}, \bibinfo {author}
  {\bibfnamefont {T.}~\bibnamefont {Pfau}}, \ and\ \bibinfo {author}
  {\bibfnamefont {H.}~\bibnamefont {Giessen}},\ }\href {\doibase
  10.1038/nmat2495} {\bibfield  {journal} {\bibinfo  {journal} {Nature Mater.}\
  }\textbf {\bibinfo {volume} {8}},\ \bibinfo {pages} {758} (\bibinfo {year}
  {2009})}\BibitemShut {NoStop}%
\bibitem [{\citenamefont {Bozhevolnyi}\ \emph {et~al.}(2011)\citenamefont
  {Bozhevolnyi}, \citenamefont {Evlyukhin}, \citenamefont {Pors}, \citenamefont
  {Nielsen}, \citenamefont {Willatzen},\ and\ \citenamefont
  {Albrektsen}}]{Bozhevolnyi2011}%
  \BibitemOpen
  \bibfield  {author} {\bibinfo {author} {\bibfnamefont {S.~I.}\ \bibnamefont
  {Bozhevolnyi}}, \bibinfo {author} {\bibfnamefont {A.~B.}\ \bibnamefont
  {Evlyukhin}}, \bibinfo {author} {\bibfnamefont {A.}~\bibnamefont {Pors}},
  \bibinfo {author} {\bibfnamefont {M.~G.}\ \bibnamefont {Nielsen}}, \bibinfo
  {author} {\bibfnamefont {M.}~\bibnamefont {Willatzen}}, \ and\ \bibinfo
  {author} {\bibfnamefont {O.}~\bibnamefont {Albrektsen}},\ }\href {\doibase
  10.1088/1367-2630/13/2/023034} {\bibfield  {journal} {\bibinfo  {journal}
  {New Journal of Physics}\ }\textbf {\bibinfo {volume} {13}},\ \bibinfo
  {pages} {023034} (\bibinfo {year} {2011})}\BibitemShut {NoStop}%
\bibitem [{\citenamefont {Liu}\ \emph {et~al.}(2011)\citenamefont {Liu},
  \citenamefont {Hentschel}, \citenamefont {Weiss}, \citenamefont
  {Alivisatos},\ and\ \citenamefont {Giessen}}]{Liu11}%
  \BibitemOpen
  \bibfield  {author} {\bibinfo {author} {\bibfnamefont {N.}~\bibnamefont
  {Liu}}, \bibinfo {author} {\bibfnamefont {M.}~\bibnamefont {Hentschel}},
  \bibinfo {author} {\bibfnamefont {T.}~\bibnamefont {Weiss}}, \bibinfo
  {author} {\bibfnamefont {A.~P.}\ \bibnamefont {Alivisatos}}, \ and\ \bibinfo
  {author} {\bibfnamefont {H.}~\bibnamefont {Giessen}},\ }\href {\doibase
  10.1126/science.1199958} {\bibfield  {journal} {\bibinfo  {journal}
  {Science}\ }\textbf {\bibinfo {volume} {332}},\ \bibinfo {pages} {1407}
  (\bibinfo {year} {2011})}\BibitemShut {NoStop}%
\bibitem [{\citenamefont {Grahn}\ \emph
  {et~al.}(2013{\natexlab{b}})\citenamefont {Grahn}, \citenamefont
  {Shevchenko},\ and\ \citenamefont {Kaivola}}]{Interfero}%
  \BibitemOpen
  \bibfield  {author} {\bibinfo {author} {\bibfnamefont {P.}~\bibnamefont
  {Grahn}}, \bibinfo {author} {\bibfnamefont {A.}~\bibnamefont {Shevchenko}}, \
  and\ \bibinfo {author} {\bibfnamefont {M.}~\bibnamefont {Kaivola}},\ }\href
  {http://arxiv.org/abs/1303.6432} {\bibfield  {journal} {\bibinfo  {journal}
  {arXiv:1303.6432}\ } (\bibinfo {year} {2013}{\natexlab{b}})}\BibitemShut
  {NoStop}%
\bibitem [{\citenamefont {Alaee}\ \emph {et~al.}(2013)\citenamefont {Alaee},
  \citenamefont {Menzel}, \citenamefont {Banas}, \citenamefont {Banas},
  \citenamefont {Xu}, \citenamefont {Chen}, \citenamefont {Moser},
  \citenamefont {Lederer},\ and\ \citenamefont {Rockstuhl}}]{Alaee13}%
  \BibitemOpen
  \bibfield  {author} {\bibinfo {author} {\bibfnamefont {R.}~\bibnamefont
  {Alaee}}, \bibinfo {author} {\bibfnamefont {C.}~\bibnamefont {Menzel}},
  \bibinfo {author} {\bibfnamefont {A.}~\bibnamefont {Banas}}, \bibinfo
  {author} {\bibfnamefont {K.}~\bibnamefont {Banas}}, \bibinfo {author}
  {\bibfnamefont {S.}~\bibnamefont {Xu}}, \bibinfo {author} {\bibfnamefont
  {H.}~\bibnamefont {Chen}}, \bibinfo {author} {\bibfnamefont {H.~O.}\
  \bibnamefont {Moser}}, \bibinfo {author} {\bibfnamefont {F.}~\bibnamefont
  {Lederer}}, \ and\ \bibinfo {author} {\bibfnamefont {C.}~\bibnamefont
  {Rockstuhl}},\ }\href {\doibase 10.1103/PhysRevB.87.075110} {\bibfield
  {journal} {\bibinfo  {journal} {Phys. Rev. B}\ }\textbf {\bibinfo {volume}
  {87}},\ \bibinfo {pages} {075110} (\bibinfo {year} {2013})}\BibitemShut
  {NoStop}%
\bibitem [{\citenamefont {Chen}(2012)}]{PA12}%
  \BibitemOpen
  \bibfield  {author} {\bibinfo {author} {\bibfnamefont {H.-T.}\ \bibnamefont
  {Chen}},\ }\href {\doibase 10.1364/OE.20.007165} {\bibfield  {journal}
  {\bibinfo  {journal} {Opt. Express}\ }\textbf {\bibinfo {volume} {20}},\
  \bibinfo {pages} {7165} (\bibinfo {year} {2012})}\BibitemShut {NoStop}%
\bibitem [{\citenamefont {Saleh}\ and\ \citenamefont
  {Teich}(2007)}]{Photonics}%
  \BibitemOpen
  \bibfield  {author} {\bibinfo {author} {\bibfnamefont {B.~E.~A.}\
  \bibnamefont {Saleh}}\ and\ \bibinfo {author} {\bibfnamefont {M.~C.}\
  \bibnamefont {Teich}},\ }\href@noop {} {\emph {\bibinfo {title} {Fundamentals
  of photonics}}},\ \bibinfo {edition} {2nd}\ ed.\ (\bibinfo  {publisher}
  {Wiley},\ \bibinfo {address} {New Jersey},\ \bibinfo {year}
  {2007})\BibitemShut {NoStop}%
\bibitem [{\citenamefont {Graham}\ and\ \citenamefont
  {Raab}(1996)}]{Graham1996}%
  \BibitemOpen
  \bibfield  {author} {\bibinfo {author} {\bibfnamefont {E.~B.}\ \bibnamefont
  {Graham}}\ and\ \bibinfo {author} {\bibfnamefont {R.~E.}\ \bibnamefont
  {Raab}},\ }\href {\doibase 10.1364/JOSAA.13.001239} {\bibfield  {journal}
  {\bibinfo  {journal} {J. Opt. Soc. Am. A}\ }\textbf {\bibinfo {volume}
  {13}},\ \bibinfo {pages} {1239} (\bibinfo {year} {1996})}\BibitemShut
  {NoStop}%
\bibitem [{\citenamefont {Graham}\ and\ \citenamefont
  {Raab}(2000)}]{Graham2000}%
  \BibitemOpen
  \bibfield  {author} {\bibinfo {author} {\bibfnamefont {E.~B.}\ \bibnamefont
  {Graham}}\ and\ \bibinfo {author} {\bibfnamefont {R.~E.}\ \bibnamefont
  {Raab}},\ }\href {\doibase 10.1098/rspa.2000.0559} {\bibfield  {journal}
  {\bibinfo  {journal} {Proc. R. Soc. Lond. A}\ }\textbf {\bibinfo {volume}
  {456}},\ \bibinfo {pages} {1193} (\bibinfo {year} {2000})}\BibitemShut
  {NoStop}%
\bibitem [{\citenamefont {Grahn}\ \emph
  {et~al.}(2012{\natexlab{a}})\citenamefont {Grahn}, \citenamefont
  {Shevchenko},\ and\ \citenamefont {Kaivola}}]{Multipole}%
  \BibitemOpen
  \bibfield  {author} {\bibinfo {author} {\bibfnamefont {P.}~\bibnamefont
  {Grahn}}, \bibinfo {author} {\bibfnamefont {A.}~\bibnamefont {Shevchenko}}, \
  and\ \bibinfo {author} {\bibfnamefont {M.}~\bibnamefont {Kaivola}},\ }\href
  {\doibase 10.1088/1367-2630/14/9/093033} {\bibfield  {journal} {\bibinfo
  {journal} {New Journal of Physics}\ }\textbf {\bibinfo {volume} {14}},\
  \bibinfo {pages} {093033} (\bibinfo {year} {2012}{\natexlab{a}})}\BibitemShut
  {NoStop}%
\bibitem [{\citenamefont {Dudley}(1994)}]{DudleyBook}%
  \BibitemOpen
  \bibfield  {author} {\bibinfo {author} {\bibfnamefont {D.~G.}\ \bibnamefont
  {Dudley}},\ }\href@noop {} {\emph {\bibinfo {title} {Mathematical foundations
  for electromagnetic theory}}}\ (\bibinfo  {publisher} {IEEE Press},\ \bibinfo
  {address} {New York},\ \bibinfo {year} {1994})\BibitemShut {NoStop}%
\bibitem [{\citenamefont {Curto}\ \emph {et~al.}(2013)\citenamefont {Curto},
  \citenamefont {Taminiau}, \citenamefont {Volpe}, \citenamefont {Kreuzer},
  \citenamefont {Quidant},\ and\ \citenamefont {van Hulst}}]{Curto13}%
  \BibitemOpen
  \bibfield  {author} {\bibinfo {author} {\bibfnamefont {A.~G.}\ \bibnamefont
  {Curto}}, \bibinfo {author} {\bibfnamefont {T.~H.}\ \bibnamefont {Taminiau}},
  \bibinfo {author} {\bibfnamefont {G.}~\bibnamefont {Volpe}}, \bibinfo
  {author} {\bibfnamefont {M.~P.}\ \bibnamefont {Kreuzer}}, \bibinfo {author}
  {\bibfnamefont {R.}~\bibnamefont {Quidant}}, \ and\ \bibinfo {author}
  {\bibfnamefont {N.~F.}\ \bibnamefont {van Hulst}},\ }\href {\doibase
  10.1038/ncomms2769} {\bibfield  {journal} {\bibinfo  {journal} {Nat.
  Commun.}\ }\textbf {\bibinfo {volume} {4}},\ \bibinfo {pages} {1750}
  (\bibinfo {year} {2013})}\BibitemShut {NoStop}%
\bibitem [{\citenamefont {Kaelberer}\ \emph {et~al.}(2010)\citenamefont
  {Kaelberer}, \citenamefont {Fedotov}, \citenamefont {Papasimakis},
  \citenamefont {Tsai},\ and\ \citenamefont {Zheludev}}]{Kaelberer10122010}%
  \BibitemOpen
  \bibfield  {author} {\bibinfo {author} {\bibfnamefont {T.}~\bibnamefont
  {Kaelberer}}, \bibinfo {author} {\bibfnamefont {V.~A.}\ \bibnamefont
  {Fedotov}}, \bibinfo {author} {\bibfnamefont {N.}~\bibnamefont
  {Papasimakis}}, \bibinfo {author} {\bibfnamefont {D.~P.}\ \bibnamefont
  {Tsai}}, \ and\ \bibinfo {author} {\bibfnamefont {N.~I.}\ \bibnamefont
  {Zheludev}},\ }\href {\doibase 10.1126/science.1197172} {\bibfield  {journal}
  {\bibinfo  {journal} {Science}\ }\textbf {\bibinfo {volume} {330}},\ \bibinfo
  {pages} {1510} (\bibinfo {year} {2010})}\BibitemShut {NoStop}%
\bibitem [{\citenamefont {Zeng}\ \emph {et~al.}(2013)\citenamefont {Zeng},
  \citenamefont {Chen},\ and\ \citenamefont {Dalvit}}]{Zeng13}%
  \BibitemOpen
  \bibfield  {author} {\bibinfo {author} {\bibfnamefont {Y.}~\bibnamefont
  {Zeng}}, \bibinfo {author} {\bibfnamefont {H.-T.}\ \bibnamefont {Chen}}, \
  and\ \bibinfo {author} {\bibfnamefont {D.~A.~R.}\ \bibnamefont {Dalvit}},\
  }\href {\doibase 10.1364/OE.21.003540} {\bibfield  {journal} {\bibinfo
  {journal} {Opt. Express}\ }\textbf {\bibinfo {volume} {21}},\ \bibinfo
  {pages} {3540} (\bibinfo {year} {2013})}\BibitemShut {NoStop}%
\bibitem [{\citenamefont {Grahn}\ \emph
  {et~al.}(2012{\natexlab{b}})\citenamefont {Grahn}, \citenamefont
  {Shevchenko},\ and\ \citenamefont {Kaivola}}]{MD1}%
  \BibitemOpen
  \bibfield  {author} {\bibinfo {author} {\bibfnamefont {P.}~\bibnamefont
  {Grahn}}, \bibinfo {author} {\bibfnamefont {A.}~\bibnamefont {Shevchenko}}, \
  and\ \bibinfo {author} {\bibfnamefont {M.}~\bibnamefont {Kaivola}},\ }\href
  {\doibase 10.1103/PhysRevB.86.035419} {\bibfield  {journal} {\bibinfo
  {journal} {Phys. Rev. B}\ }\textbf {\bibinfo {volume} {86}},\ \bibinfo
  {pages} {035419} (\bibinfo {year} {2012}{\natexlab{b}})}\BibitemShut
  {NoStop}%
\bibitem [{\citenamefont {Johnson}\ and\ \citenamefont
  {Christy}(1972)}]{Johnson1972}%
  \BibitemOpen
  \bibfield  {author} {\bibinfo {author} {\bibfnamefont {P.~B.}\ \bibnamefont
  {Johnson}}\ and\ \bibinfo {author} {\bibfnamefont {R.~W.}\ \bibnamefont
  {Christy}},\ }\href {\doibase 10.1103/PhysRevB.6.4370} {\bibfield  {journal}
  {\bibinfo  {journal} {Phys. Rev. B}\ }\textbf {\bibinfo {volume} {6}},\
  \bibinfo {pages} {4370} (\bibinfo {year} {1972})}\BibitemShut {NoStop}%
\end{thebibliography}

% Data from .bbl file:
%---------------------------------------------------------------------------------
%merlin.mbs apsrev4-1.bst 2010-07-25 4.21a (PWD, AO, DPC) hacked
%Control: key (0)
%Control: author (8) initials jnrlst
%Control: editor formatted (1) identically to author
%Control: production of article title (-1) disabled
%Control: page (0) single
%Control: year (1) truncated
%Control: production of eprint (0) enabled
%

%---------------------------------------------------------------------------------

% The contents and all included figures belong to the authors
% Patrick Grahn, Andriy Shevchenko and Matti Kaivola

\end{document}